\documentclass[floatfix,aps,amsmath,nofootinbib,onecolumn,10pt]{revtex4}

\usepackage{graphicx}
\usepackage{bm}
\usepackage{rotating}
\usepackage{array}

\def\({\left(}
\def\){\right)}
\def\[{\left[}
\def\]{\right]}

\def\e{\begin{equation}}
\def\q{\end{equation}}
\def\m{\begin{eqnarray}}
\def\n{\end{eqnarray}}

\begin{document}

\title{New cosmological constraints with extended-Baryon Oscillation Spectroscopic Survey DR14 quasar sample}

\author{Lu Chen$^{1,2}$ \footnote{chenlu@itp.ac.cn}, Qing-Guo Huang$^{1,2,3}$ \footnote{huangqg@itp.ac.cn} and Ke Wang$^{1,2}$ \footnote{wangke@itp.ac.cn}}
\affiliation{$^1$ CAS Key Laboratory of Theoretical Physics,\\ Institute of Theoretical Physics, \\Chinese Academy of Sciences, Beijing 100190, China\\
$^2$ School of Physical Sciences, \\University of Chinese Academy of Sciences,\\ No. 19A Yuquan Road, Beijing 100049, China\\
$^3$ Synergetic Innovation Center for Quantum Effects and Applications, Hunan Normal University, Changsha 410081, China}

\date{\today}

\begin{abstract}

We update the constraints on the cosmological parameters by adopting the Planck data released in 2015 and Baryon Acoustic Oscillation (BAO) measurements including the new DR14 quasar sample measurement at redshift $z=1.52$, and we conclude that the based six-parameter $\Lambda$CDM model is preferred. Exploring some extensions to the $\Lambda$CDM model, we find that the equation of state of dark energy reads $w=-1.036\pm 0.056$ in the $w$CDM model, the effective relativistic degrees of freedom in the Universe is $N_\text{eff}=3.09_{-0.20}^{+0.18}$ in the $N_\text{eff}+\Lambda$CDM model and the spatial curvature parameter is $\Omega_k=(1.8\pm1.9)\times 10^{-3}$ in the $\Omega_k+\Lambda$CDM model at $68\%$ confidence level (C.L.), and the $95\%$ C.L. upper bounds on the sum of three active neutrinos masses are $\sum m_\nu<0.16$ eV for the normal hierarchy (NH) and $\sum m_\nu<0.19$ eV for the inverted hierarchy (IH) with $\Delta\chi^2\equiv \chi^2_\text{NH}-\chi^2_\text{IH}=-1.25$.


\end{abstract}

\pacs{???}

\maketitle


\section{Introduction}

The accuracy of cosmological observations has been significantly improved in the past two decades. The based six-parameter $\Lambda$CDM model is strongly supported by the precise measurements of anisotropies of the cosmic microwave background (CMB) \cite{Komatsu:2008hk,Ade:2015xua}. The Type Ia supernova (SNe) \cite{Conley:2011ku,Suzuki:2011hu} and Baryon Acoustic Oscillation (BAO) \cite{Cole:2005sx,Eisenstein:2005su} as a geometric complement directly encode the information of expansion history in the late-time Universe. As an important parameter characterizing the today's expansion rate, the Hubble constant is directly measured by Hubble Space Telescope (HST) \cite{Riess:2016jrr}. 


The BAO measurement is the periodic relic of fluctuations of baryonic matter density in the Universe. It is considered as a standard ruler of the universe and can be used as an independent way to constrain models.
In the previous observations, the BAO is traced directly by galaxies at low redshift and measured indirectly by analysis of Lyman-$\alpha$ (Ly$\alpha$) forest in quasar spectra at high redshifts.
Recently, the extended-Baryon Oscillation Spectroscopic Survey (eBOSS) \cite{Dawson:2015wdb} released their another percent level BAO measurement at $z=1.52$ using the auto-correlation of quasars directly, referred to as DR14 quasar sample \cite{Ata:2017dya}. It is a new method to achieve BAO features, which makes DR14 as the first BAO distance observations in the range of $1<z<2$.

The higher redshift at which BAO is measured, the more sensitive to the Hubble parameter. Therefore, we can expect an improvement in constraints on the equation of state (EOS) of dark energy (DE) and a preciser description of the expansion history by including DR14.
On the other hand, with increasing total active neutrino mass at fixed $\theta_*$, the spherically-averaged BAO distance $D_V(z\lesssim1)$ increases accordingly, but $D_V(z>1)$ falls \cite{Ade:2013zuv}. It implies that DR14 may improve the constraint on the total active neutrino mass. On the contrary, with increasing the effective number of relativistic species $N_{\textrm{eff}}$ at a fixed $\theta_*$ and a fixed redshift of matter-radiation equality $z_{\textrm{eq}}$, $D_V(z)$ decreases for all BAO measurements \cite{Ade:2013zuv}. Therefore, DR14 can improve the constraint on $N_{\textrm{eff}}$ as well.
In addition, the spatial curvature of our universe can be also constrained better because the geometry of space affects the detection of BAO measurement directly and the new released BAO measurement DR 14 fills the gap between $1<z<2$.

In this paper, we update the constraints on the EOS of DE, the active neutrino masses, the dark radiation and the spatial curvature with the Planck data and the BAO measurements including the DR14 quasar sample at $z=1.52$. The paper is arranged as follows. In Sec.~\ref{method}, we explain our methodology and the data we used. In Sec.~\ref{results}, the results for different models are presented. At last, a brief summary and discussion are included in Sec.~\ref{sum}.

\section{Data and Methodology}
\label{method}
We use the combined data of CMB and BAO measurements to constrain the parameters in the different models. Concretely, we use Planck TT,TE,EE+lowP released by the Planck Collaboration in 2015 \cite{Ade:2015xua}, namely P15, as well as the BAO measurements at $z=0.106,\ 0.15,\ 0.32,\ 0.57,\ 1.52$, namely 6dFGS \cite{Beutler:2011hx}, MGS \cite{Ross:2014qpa}, DR12 BOSS LOWZ, DR12 BOSS CMASS \cite{Cuesta:2015mqa,Gil-Marin:2015nqa}, and DR14 eBOSS quasar sample \cite{Ata:2017dya} seperately.

To show the BAO data we used, we should introduce the BAO model briefly, which is the basic model of the BAO signal. The volume-averaged values are measured, in \cite{Eisenstein:2005su}, by 
\m
D_V(z)=\[(1+z)^2   D_A^{2}(z)  \frac{cz}{H(z)}\]^{1/3}
\n
where $c$ is the light speed, $D_A(z)$ is the proper angular diameter distance \cite{Aubourg:2014yra}, given by
\m
D_A(z)=\dfrac{c}{1+z}   S_k\(\int_{0}^{z}  {  dz'\over H(z')}\)\label{con:da}
\n
where $S_k(x)$ is
\begin{eqnarray}S_k(x)=
\begin{cases}
\sin(\sqrt{-\Omega_k}x)/\sqrt{-\Omega_k},&\Omega_k<0 \\ \sinh(\sqrt{\Omega_k}x)/\sqrt{\Omega_k},&\Omega_k>0 \\x &\Omega_k=0 \
\end{cases}
\end{eqnarray}
and $H(z)$ is
\m
H(z)=H_0 \[\Omega_r (1+z)^4 + \Omega_m (1+z)^3 +\Omega_k (1+z)^2 +(1-\Omega_r -\Omega_m-\Omega_k)f(z)\]^{1/2}
\n
where 
\m
f(z)\equiv {\rho_\text{DE}(z)\over \rho_\text{DE}(0)}=\exp\[\int_o^z3(1+w(z')){dz'\over 1+z'}\], 
\n
and $w(z)\equiv p_\text{DE}/\rho_\text{DE}$ is the EOS of DE.

\section{Results}
\label{results}

In this section, we will represent our new constraints on the dark energy, the neutrino masses, the dark radiation and the spatial curvature of the universe separatelly.

\subsection{Constraints on Dark Energy}

In this subsection, we constrain the cosmological parameters in the $\Lambda$CDM model, the $w$CDM model and the $w_0 w_a$CDM model \cite{Chevallier:2000qy,Linder:2002et} respectively. Our results are summarized in Tab.~\ref{tablede}. We run CosmoMC \cite{Lewis:2002ah} in the $\Lambda$CDM model as the basic model, where there are six free cosmological parameters ${\lbrace\Omega_b h^2,\Omega_c h^2,100\theta_{MC},\tau,n_s,\ln(10^{10}A_s)\rbrace}$. Here $\Omega_b h^2$ is the density of the baryonic matter today, $\Omega_c h^2$ is the cold dark matter density today, $100\theta_{MC}$ is 100 times the ratio of the angular diameter distance to the large scale structure sound horizon, $\tau$ is the optical depth, $n_s$ is the scalar spectrum index and $A_s$ is the amplitude of the power spectrum of primordial curvature perturbations.  
\begin{table}
\caption{The 68$\%$ limits for the cosmological parameters in different DE models from P15+BAO.}\label{tablede}
\begin{tabular}{p{3 cm}<{\centering}|p{3.5cm}<{\centering} p{3.5cm}<{\centering} p{3.5cm}<{\centering} }
\hline
                & $\Lambda$CDM & $w$CDM  & $w_0 w_a$CDM      \\
\hline
$\Omega_b h^{2}$ & $0.02233\pm 0.00014$ & $0.02229\pm 0.00015$ & $0.02224\pm 0.00015$  \\

$\Omega_c h^{2}$ & $0.1186\pm 0.0010$   & $0.1191\pm 0.0013$   & $0.1200\pm 0.0014$    \\

$100\theta_{MC}$ & $1.04091\pm 0.00030 $ &$1.04086\pm 0.00031 $  & $1.04075\pm 0.00031 $  \\

$\tau$           & $0.085\pm 0.016$     & $0.082\pm 0.017$     & $0.075\pm 0.017$      \\

$\ln(10^{10}A_s)$ &$3.102\pm 0.032$& $3.098_{-0.032}^{+0.035}$& $3.086\pm 0.033 $       \\

$n_s$            & $0.9677\pm 0.0040 $ & $0.9664 \pm 0.0046 $ & $0.9640\pm 0.0046 $    \\
$w$              &  -                    &$-1.036\pm 0.056$      & -                     \\
 $w_0$           &    -                  & -                    & $-0.25\pm 0.32 $      \\
  $w_a$          &   -                  & -                    & $-2.29_{-0.91}^{+1.10}$\\
\hline
$H_0$ [km s$^{-1}$ Mpc$^{-1}$]&$67.81_{-0.46}^{+0.47}$ &$68.66_{-1.55}^{+1.41}$ &$62.56_{-2.74}^{+2.42}$\\
\hline
\end{tabular}
\end{table}



\begin{figure}[]
\begin{center}
\includegraphics[scale=0.5]{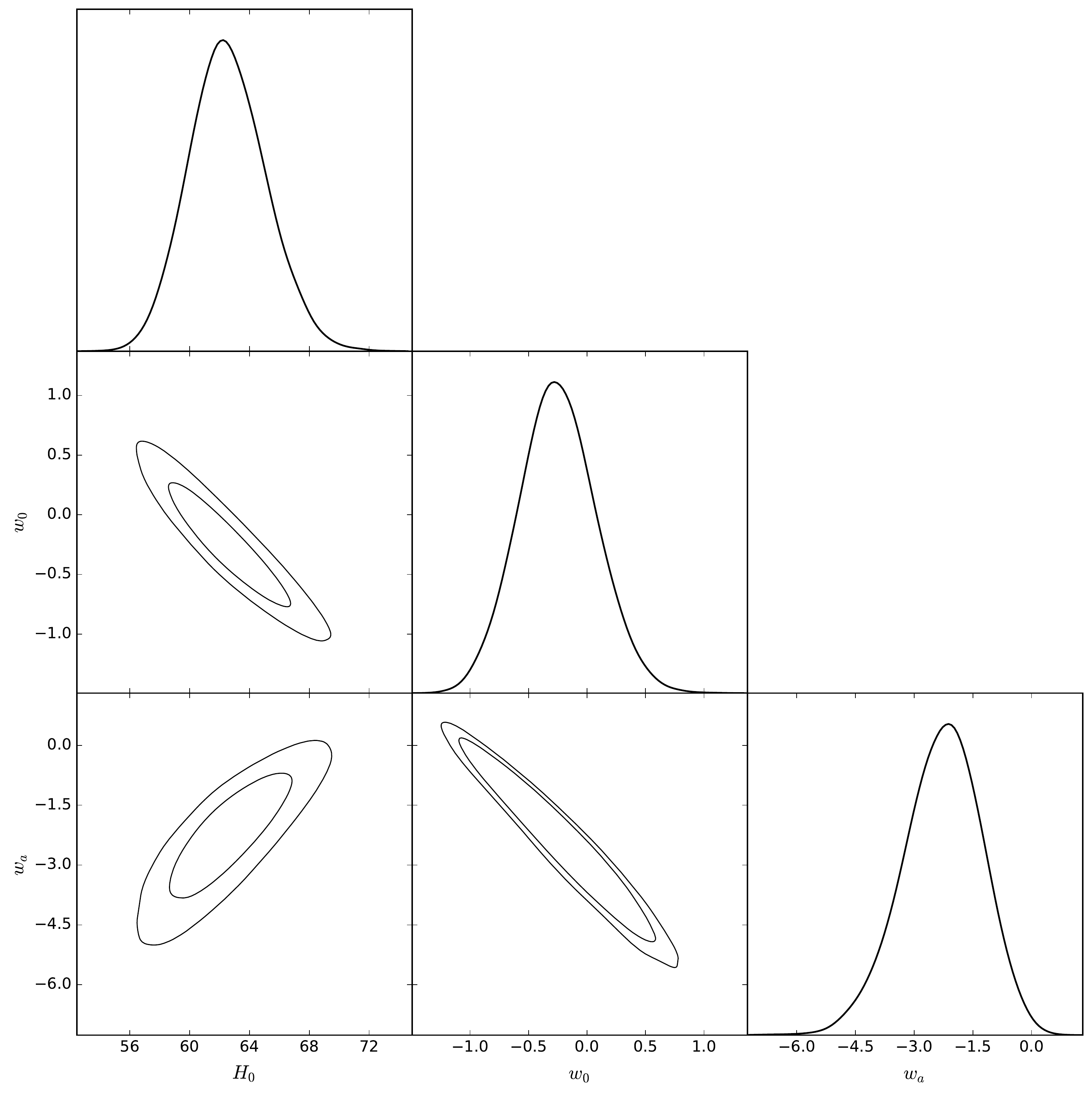}
\end{center}
\caption{The triangular plot of $H_0$, $w_0$ and $w_a$ in the $w_0 w_a$CDM model. The  point ($w_0=-1$, $w_a=0$) locates in the $w_0 w_a$CDM 68$\%$ C.L. region.}
\label{fig:ww}
\end{figure}

\begin{figure}[]
\begin{center}
\includegraphics[scale=0.9]{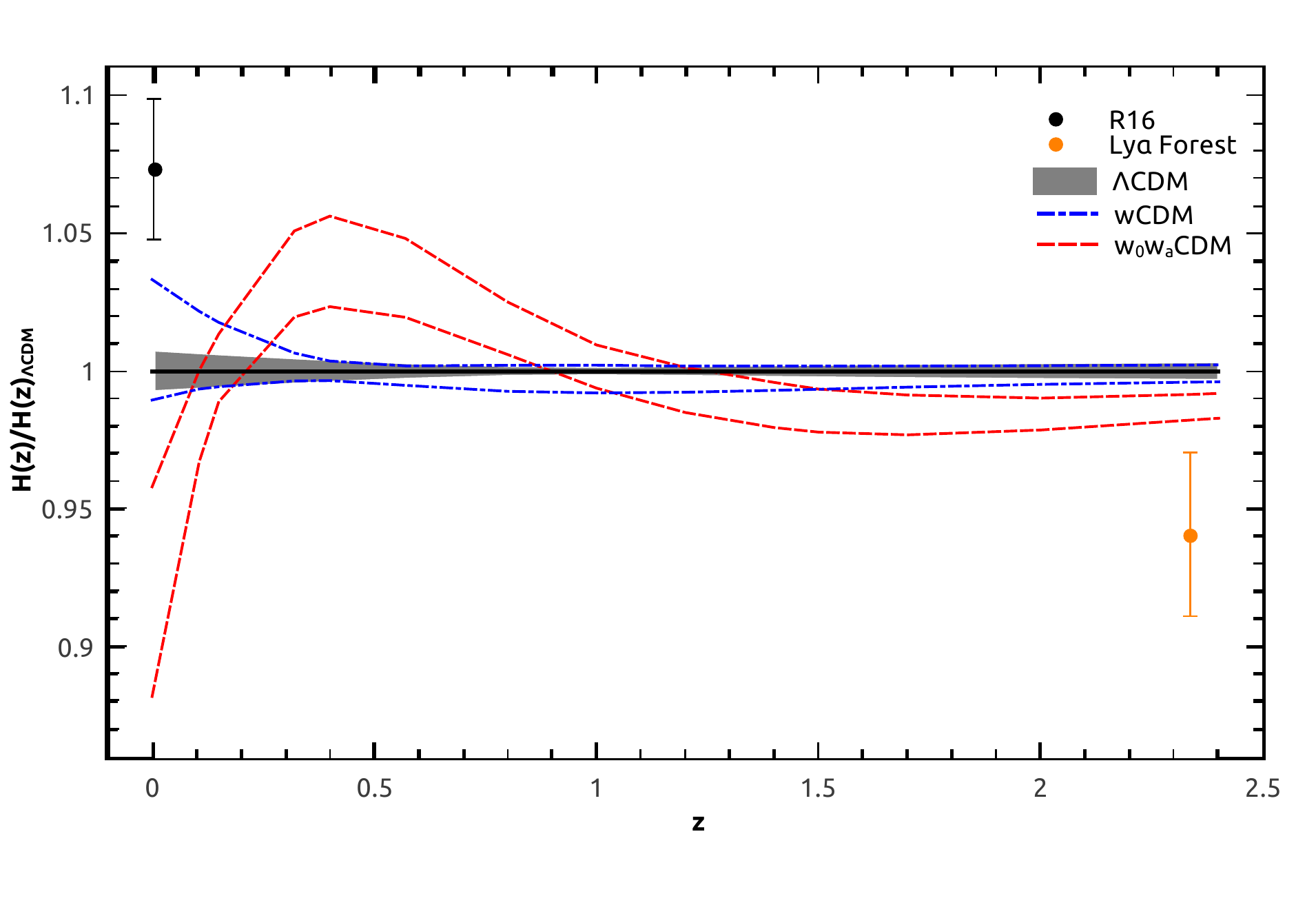}
\end{center}
\caption{The normalized $H(z)$ plot in the $\Lambda$CDM model, $w$CDM model and  $w_0 w_a$CDM model. The grey band represents the $68\%$ confidence range in the $\Lambda$CDM model allowed by P15+BAO. The ranges between the two blue dash-dotted lines and the two red dashed lines represent the $68\%$ confidence ranges of the $w$CDM model and $w_0 w_a$CDM model allowed by P15+BAO respectively. 
The black and orange error bars denote the Hubble parameters measured by HST (named R16) in \cite{Riess:2016jrr} and the Ly$\alpha$ forest of BOSS DR11 quasars (named Ly$\alpha$ Forest) in \cite{Delubac:2014aqe}. 
}
\label{fig:H}
\end{figure}

The EOS of DE is $w=-1.036\pm0.056$ in the $w$CDM model at $68\%$ confidence level (C.L.). The triangular plot of $H_0$, $w_0$ and $w_a$ in the $w_0 w_a$CDM model is shown in Fig.~\ref{fig:ww} and it indicates that the prediction of $\Lambda$CDM is within the $68\%$ confidence region in this figure, which seems to be in conflict with the $w_0$, $w_a$ values in Tab.~\ref{tablede}. Actually, the probabilities are the integrated probabilities, which means the values in the Tab.~\ref{tablede} have been marginalized over all the other parameters except the aimed parameter. Due to the strong correlation between $w_0$ and $w_a$, we should check if the prediction of $\Lambda$CDM is consistent with datasets in the $w_0-w_a$ 2D contour plot.
Marginalizing over the other cosmological parameters, we also plot the evolution of the normalized Hubble parameter $H(z)$ in Fig.~\ref{fig:H} where the Hubble parameter is normalized by comparing to those in the best-fit $\Lambda$CDM model. 
The Hubble constant is $H_0=67.81_{-0.46}^{+0.47}$ km s$^{-1}$ Mpc$^{-1}$ in the $\Lambda$CDM model, $H_0=68.66_{-1.55}^{+1.41}$ km s$^{-1}$ Mpc$^{-1}$ in the $w$CDM model, and $H_0=62.56_{-2.74}^{+2.42}$ km s$^{-1}$ Mpc$^{-1}$ in the $w_0w_a$CDM model, at 68$\%$ C.L.. In all, there is a significant tension on the measurement of Hubble constant between global fitting P15+BAO and the direct measurement by HST in \cite{Riess:2016jrr} (named R16) which gives $H_0=73.24\pm 1.74$ km s$^{-1}$ Mpc$^{-1}$. Even though such a tension is slightly relaxed in the $w$CDM model, it is aggravated in the $w_0 w_a$CDM model. In order to significantly relax such a tension, a more dramatic design of the EOS of DE is needed \cite{Qing-Guo:2016ykt}. In addition, an around $2\sigma$ tension on the Hubble parameter at $z=2.34$ between the predictions of these three DE models constrained by P15+BAO and the measurement by Ly$\alpha$ forest of BOSS DR11 quasars \cite{Delubac:2014aqe} which gives $H(z=2.34)=222\pm 7$ km s$^{-1}$ Mpc$^{-1}$ still exists.

\subsection{Constraints on the total mass of active neutrinos}

\begin{figure}[]
\begin{center}
\includegraphics[scale=0.9]{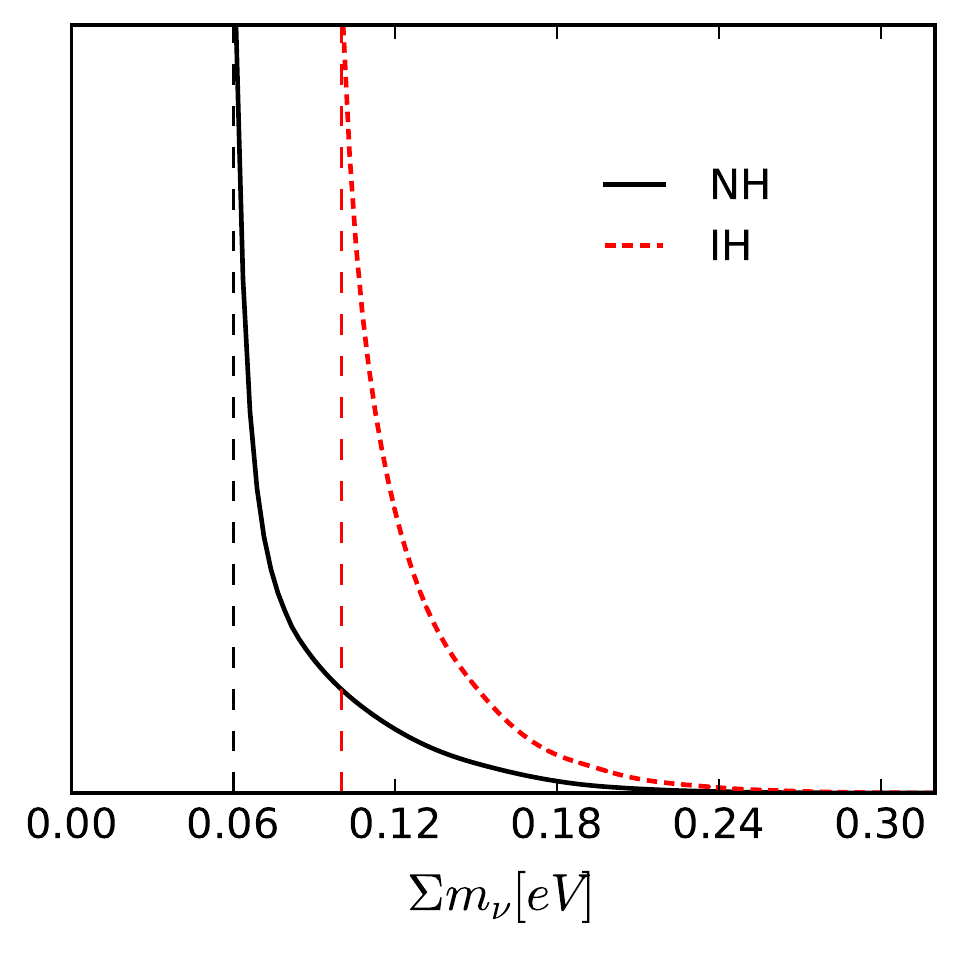}
\end{center}
\caption{The likelihood distributions of $\sum m_{\nu}$ for the NH and IH neutrinos in the $\nu \Lambda$CDM model. The dashed lines donate the allowed minimums of $\sum m_{\nu}$, namely 0.059 eV for NH and 0.101 eV for IH.}
\label{fig:mnu}
\end{figure}

The neutrino oscillation implies that the active neutrinos have mass splittings
\m
\Delta m_{21}^2=m_2^2-m_1^2,
\n
\m
 \vert \Delta m_{31}^2\vert=\vert m_3^2-m_1^2\vert.
\n
where $\Delta m_{21}^2\simeq 7.54\times 10^{-5}$ eV$^2$ and $\vert \Delta m_{31}^2\vert\simeq 2.46\times 10^{-3}$ eV$^2$ \cite{Olive:2016xmw}. That is to say, there are two possible mass hierarchies: if $m_1<m_2<m_3$, it's a normal hierarchy (NH); if $m_3<m_1<m_2$, it's an inverted hierarchy (IH). 
The neutrino mass spectrum is expressed as 
\begin{eqnarray}(m_1,m_2,m_3)=
\begin{cases}
(m_1,\sqrt{m_1^2+\Delta m_{21}^2},\sqrt{m_1^2+\vert \Delta m_{31}^2\vert}),& \text{for\ NH,\  where \ }\text{$m_1$ \ is \ the\  minimum}, \\ (\sqrt{m_3^2+\vert \Delta m_{31}^2\vert},\sqrt{m_3^2+\Delta m_{21}^2+\vert \Delta m_{31}^2\vert},m_3),& \text{for \ IH,\ where \ $m_3$ \ is \ the\  minimum},\\
\end{cases}
\end{eqnarray}
and the total mass satisfies 
\begin{eqnarray}\sum m_{\nu}=m_1+m_2+m_3\gtrsim
\begin{cases}
0.059,& \text{for\ NH}, \\ 0.101,& \text{for \ IH.}\\
\end{cases}
\end{eqnarray}
Here we set the minimum of the three neutrino masses as a free parameter and the sum of the neutrino masses as a derived parameter. Our results are summarized in Tab.~\ref{tab:neutrino}. 
\begin{table}
\caption{The 68$\%$ limits for the cosmological parameters in the $\nu_\text{NH}\Lambda$CDM model and the $\nu_\text{IH}\Lambda$CDM model from P15+BAO. }
\begin{tabular}{p{3 cm}<{\centering}|p{3.5cm}<{\centering} p{3.5cm}<{\centering} p{3.5cm}<{\centering}  }
\hline
\                       &   $\nu_\text{NH}\Lambda$CDM &$\nu_\text{IH}\Lambda$CDM       \\
\hline
$\Omega_b h^{2}$        &   $0.02234\pm 0.00014$   &$0.02235\pm 0.00014$           \\
$\Omega_c h^{2}$        &   $0.1184\pm 0.0011$     &$0.1181\pm 0.0010$             \\
$100\theta_{MC}$        &     $1.04093\pm 0.00029 $ &$1.04093\pm 0.00029 $            \\
$\tau$                  &   $0.087\pm 0.0165$       &$0.090_{-0.016}^{+0.018}$               \\
$\ln(10^{10}A_s)$       &    $3.106\pm 0.033 $     &$3.111_{-0.032}^{+0.035} $              \\
$n_s$                   &   $0.9682^{+0.0041}_{-0.0040} $   &$0.9689\pm 0.0041$  \\
$m_{\nu,min}$ ($95\%$ C.L.)    &   $<$   0.047 eV         &  $<$ 0.049 eV                \\
$\Sigma m_{\nu}$ ($95\%$ C.L.) &  $<$ 0.16 eV              &$<$ 0.19 eV                    \\
\hline
$H_0$ [km s$^{-1}$ Mpc$^{-1}$]&$67.65\pm{0.50}$ &$67.43\pm 0.48$ \\
\hline
\end{tabular}
\label{tab:neutrino}
\end{table}
The likelihood distribution of $\sum m_\nu$ for the NH and IH are illustrated in Fig.~\ref{fig:mnu}. 

In summary, the masses of the lightest neutrinos in NH and IH are $m_{\nu,min}<0.047$ eV and $m_{\nu,min}<0.049$ eV at 95$\%$ C.L. respectively. The total active neutrino masses are given by $\sum m_{\nu}<0.16$ eV and $\sum m_{\nu}<0.19$ eV for the NH and IH, and the NH is slightly preferred with $\Delta\chi^2\equiv \chi^2_\text{NH}-\chi^2_\text{IH}=-1.25$. Our new results are slightly tighter than those without the DR14 quasar sample \cite{Huang:2015wrx}. See some other related investigations in \cite{Xu:2016ddc,Guo:2017hea,Li:2017iur,Capozzi:2017ipn,Feng:2017nss,Feng:2017mfs,Wang:2017htc,Vagnozzi:2017ovm,Giusarma:2016phn}. 



\subsection{Constraints on the dark radiation}

The total energy density of radiation in the Universe is given by 
\m
\rho_r=\[1+\dfrac{7}{8}(\dfrac{4}{11})^{4/3} N_\text{eff} \] \rho_\gamma
\n
where $\rho_\gamma$ is the CMB photon energy density, $N_\text{eff}$ denotes the effective relativistic degrees of the freedom in the Universe. For the three standard model neutrinos, their contribution to $N_\text{eff}$ is 3.046 due to non-instantaneous decoupling corrections. Then the additional relativistic degree of freedom $\Delta N_\text{eff}\equiv N_\text{eff}-3.046$ implies the existence of some other unknown sources of relativistic degree of freedom. $\Delta N_\text{eff}<0$ is considered to result from incompletely thermalized neutrinos or the existence of photons produced after neutrino decoupling, which is less motivated. But there exists many cases that $\Delta N_\text{eff}>0$. If a kind of additional massless particles don't interact with others since the epoch of recombination, their energy density evolves exactly like radiation and thus contributes $\Delta N_\text{eff}=1$. There are more explanation for $0<\Delta N_\text{eff}<1$ considering the non-thermal case and the bosonic particles. The thermalized massless boson decoupled during 0.5 MeV$<T<100$MeV contributes $\Delta N_\text{eff}\simeq 0.57$ and $\Delta N_\text{eff}\simeq 0.39$ if they decoupled before $T= 100$ MeV \cite{Weinberg:2013kea}.

\begin{figure}[]
\centering
\includegraphics[scale=0.9]{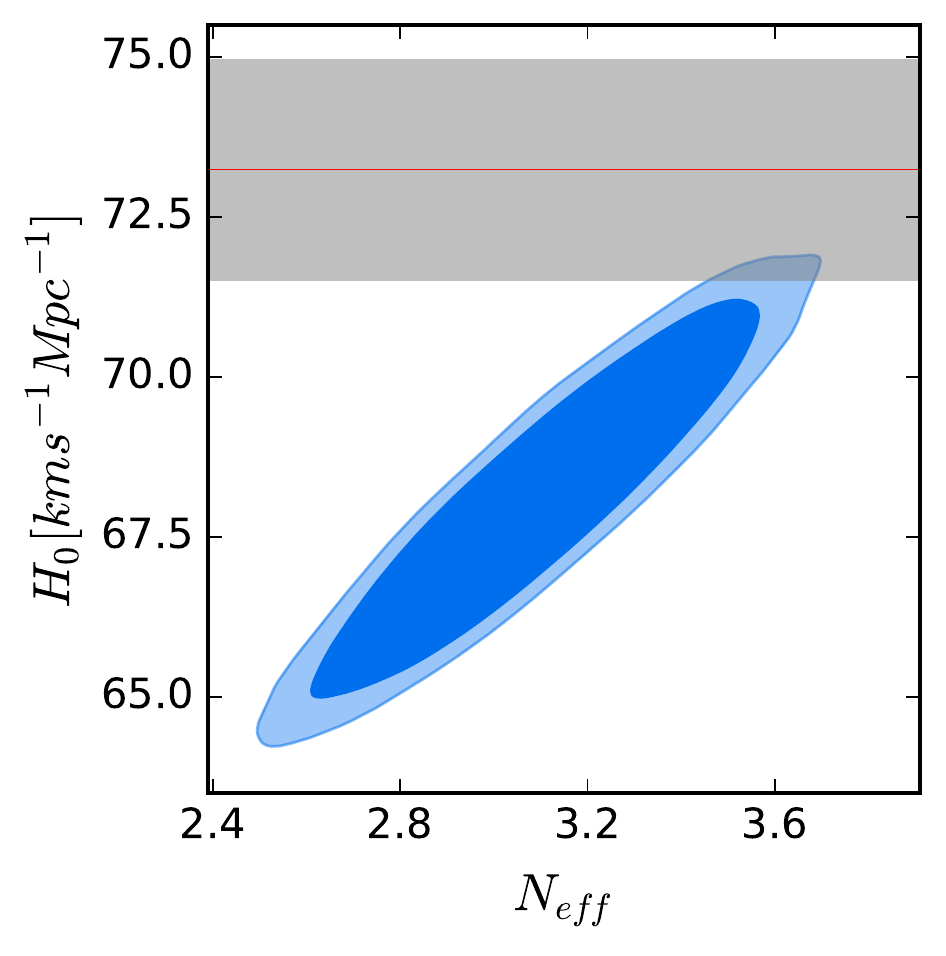}
\caption{The 2D contour plot of $H_0$ and $N_\text{eff}$ in the $N_\text{eff}+\Lambda$CDM model. The rectangle shaded region represents the observational value $H_0=73.24\pm 1.74$ km s$^{-1}$ Mpc$^{-1}$. They are overlapped at the 95$\%$ C.L. region.}\label{fig:nnu}
\end{figure}

In the $N_\text{eff}+\Lambda $CDM model, $N_\text{eff}$ is taken as a free parameter. The results are summarized in Tab.~\ref{paramsnnu}.
\begin{table}
\caption{The 68$\%$ limits for the cosmological parameters in the $N_\text{eff}+\Lambda $CDM model from P15+BAO.}\label{paramsnnu}
\begin{tabular}{p{3 cm}<{\centering}|p{3.5cm}<{\centering} p{3.5cm}<{\centering} p{3.5cm}<{\centering}  }
\hline
\                       & $N_\textsf{eff}+\Lambda$CDM   \\
\hline
$\Omega_b h^{2}$        &$0.02236\pm 0.00019$  \\
$\Omega_c h^{2}$        &$0.1193\pm 0.0031$   \\
$100\theta_{MC}$        &$1.04085\pm 0.00044 $  \\
$\tau$                  &$0.086\pm 0.017$ \\
$\ln(10^{10}A_s)$       &$3.105\pm 0.035$  \\
$n_s$                   &$0.9691_{-0.0075}^{+0.0076}$\\
$N_\textsf{eff} $       &$3.09_{-0.20}^{+0.18}$\\
\hline
$H_0$ [km s$^{-1}$ Mpc$^{-1}$]&$68.07_{-1.20}^{+1.21}$\\
\hline
\end{tabular}
\end{table}
Our results give $N_\text{eff}=3.09_{-0.20}^{+0.18}$ at 68$\%$ C.L., which is consistent with the fact that there are only three active neutrinos in the Universe. 
On the other hand, for example in \cite{Riess:2016jrr}, the dark radiation is proposed to relax the tension on the Hubble constant between the global fitting P15+BAO and the direct measurement by HST. Here we illustrate the constraints on $H_0$ and $N_\text{eff}$ in Fig.~\ref{fig:nnu}. 
From Fig.~\ref{fig:nnu}, we find that the dark radiation cannot really solve this tension. 


\subsection{Constraints on the curvature}

\begin{figure}[]
\centering
\includegraphics[scale=0.9]{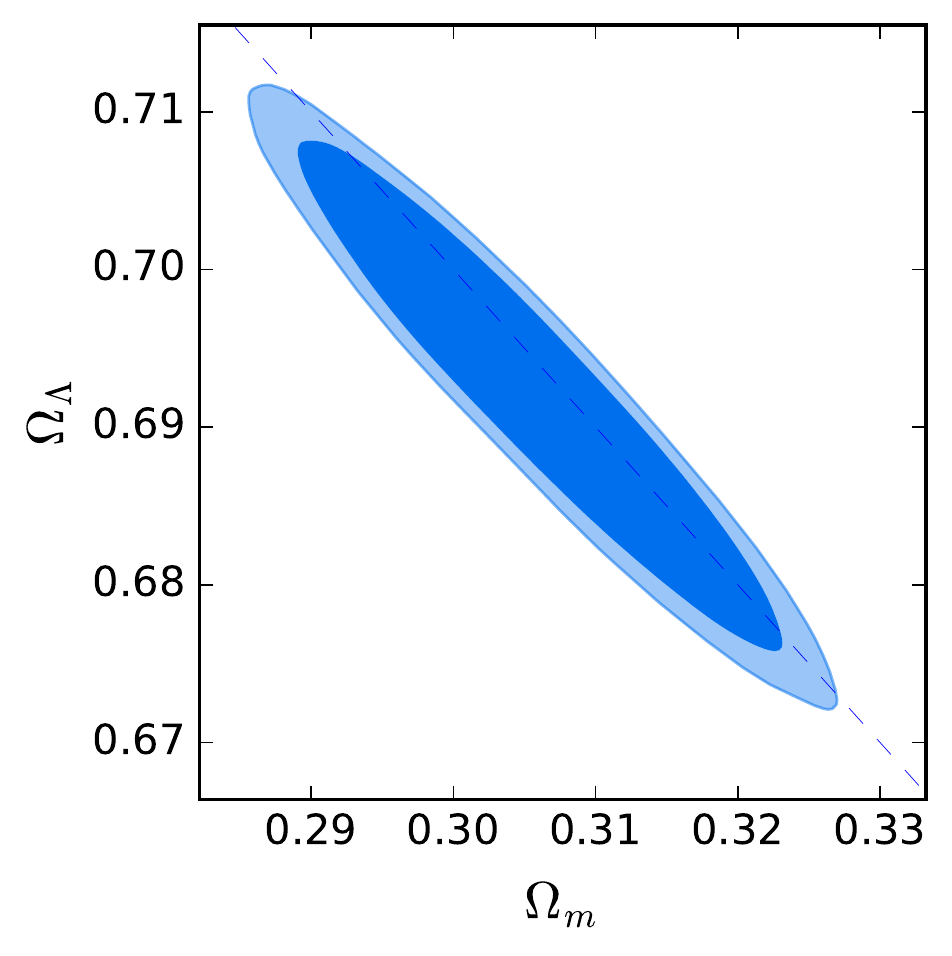}
\caption{The contour plot of $\Omega_m$ and $\Omega_{\Lambda}$ in the $\Omega_k+\Lambda$CDM model. The dashed line indicates $\Omega_m+\Omega_{\Lambda}=1$.}
\label{fig:omegak}
\end{figure}

According to Eq.~(\ref{con:da}), the spatial geometry affects the distance measurements, and hence the spatial curvature parameter $\Omega_k$ can be constrained by using BAO data. In the $\Omega_k$+$\Lambda$CDM model, $\Omega_k$ is taken as a free parameter. The constraints on the cosmological parameters in the $\Omega_k$+$\Lambda$CDM model are given in Tab.~\ref{paramsomegak}. 
\begin{table}
\centering
\caption{The $68\%$ limits for the cosmological parameters in the $\Omega_k+\Lambda$CDM model from P15+BAO }\label{paramsomegak}
\begin{tabular}{p{3 cm}<{\centering}| p{5cm}<{\centering}}
\hline
 \  &  $\Omega_k$+$\Lambda$CDM \\
\hline
$\Omega_b h^{2}$ &$0.02226\pm 0.00016$ \\
$\Omega_c h^{2}$ &$0.1196\pm 0.0015$ \\
$100\theta_{MC}$ &$1.04078\pm 0.00030 $\\
$\tau$           &$0.081\pm 0.017$\\
$\ln(10^{10}A_s)$ &$3.097\pm 0.033 $\\
$n_s$            &$0.9652 \pm 0.0048 $\\
$\Omega_k$ &$(1.8\pm 1.9)\times10^{-3}$\\
\hline
$H_0$ [km s$^{-1}$ Mpc$^{-1}$]&$68.27_{-0.95}^{+0.68}$ \\
\hline
\end{tabular}
\end{table}
We find that the spatial curvature has been tightly constrained, namely $\Omega_k=(1.8\pm 1.9)\times10^{-3}$ at $68\%$ C.L. and $\Omega_k=(1.8_{-3.8}^{+3.9})\times10^{-3}$ at $95\%$ C.L. which is nicely consistent with a spatially flat universe. Adopting P15 only, the constraint on the spatial curvature is $\Omega_k=(-40_{-41}^{+38})\times10^{-3}$ at $95\%$ C.L. which is around one oder of magnitude looser comparing to our new result. However, our results improves little comparing the Planck $+$ BAO result in the Planck table, $\Omega_k=(0.2\pm 2.1)\times10^{-3}$ at $68\%$ C.L., which implies that the DR14 sample helps little to constrain the curvature. The constraints on $\Omega_\Lambda$ and $\Omega_m$ are illustrated in Fig.~\ref{fig:omegak}. 


\section{Summary and discussion}
\label{sum}

In this paper we provide the new constraints on the cosmological parameters in some extensions to the based six-parameter $\Lambda$CDM model by combining P15 and BAO data including the DR14 quasar sample measurement released recently by eBOSS. We do not find any signals beyond this based cosmological model. 

We explore the EOS of DE in two extended models, namely $w$CDM and $w_0 w_a$CDM model, and find $w=-1.036\pm 0.056$ at $68\%$ C.L. in the $w$CDM model, $w_0=-0.25 \pm 0.32$, $w_a=-2.29^{+1.10}_{-0.91}$ at $68\%$ C.L. in the $w_0 w_a$CDM model and $w=-1$ is located within the $68\%$ C.L. region. But the tension on the Hubble constant with the direct measurement by HST and the global fitting P15+BAO in $w$CDM model cannot be significantly relaxed and the $w_0 w_a$CDM model makes even worse. The neutrino mass normal hierarchy is slightly preferred by $\Delta\chi^2\equiv \chi^2_\text{NH}-\chi^2_\text{IH}=-1.25$ compared to the inverted hierarchy, and the $95\%$ C.L. upper bounds on the sum of three active neutrinos masses are $\sum m_\nu<0.16$ eV for the normal hierarchy and $\sum m_\nu<0.19$ eV for the inverted hierarchy. The three active neutrinos are nicely consistent with the constraint on the effective relativistic degrees of freedom with $N_\text{eff}=3.09_{-0.20}^{+0.18}$ at $68\%$ C.L., and a spatially flat Universe is preferred.




\vspace{5mm}
\noindent {\bf Acknowledgments}

We acknowledge the use of HPC Cluster of ITP-CAS.
This work is supported by grants from NSFC (grant NO. 11335012, 11575271, 11690021), Top-Notch Young Talents Program of China, and partly supported by Key Research Program of Frontier Sciences, CAS.



\end{document}